# A Reliability-based Framework for Multi-path Routing Analysis in Mobile Ad-Hoc Networks

## Marcello Caleffi †, Giancarlo Ferraiuolo *, Luigi Paura †•

† Dipartimento di Ingegneria Elettronica e delle Telecomunicazioni (DIET) - Università degli Studi di Napoli Federico II – Napoli, ITALY
• Laboratorio Nazionale di Comunicazioni Multimediali (CNIT) – Napoli, ITALY
* Autorità per le Garanzie nelle Comunicazioni (AGCOM) - Centro Direzionale Isola B5 Torre Francesco - 80143 Napoli, ITALY

E-mail addresses: {name.surname}@unina.it

**Abstract:** Unlike traditional routing procedures that, at the best, single out a unique route, multi-path routing protocols discover proactively several alternative routes. It has been recognized that *multi-path routing* can be more efficient than traditional one mainly for mobile ad hoc networks, where route failure events are frequent. Most studies in the area of multi-path routing focus on heuristic methods, and the performances of these strategies are commonly evaluated by numerical simulations. The need of a theoretical analysis motivates such a paper, which proposes to resort to the terminal-pair routing reliability as performance metric. This metric allows one to assess the performance gain due to the availability of route diversity. By resorting to graph theory, we propose an analytical framework to evaluate the tolerance of multi-path route discovery processes against route failures for mobile ad hoc networks. Moreover, we derive a useful bound to easily estimate the performance improvements achieved by multi-path routing with respect to any traditional routing protocol. Finally, numerical simulation results show the effectiveness of this performance analysis.

**Keywords**: MANET; reliability; multi-path; routing; ad-hoc; graph theory; overlay graph,

**Biographical notes:**

Marcello Caleffi was born in Bondeno, Italy, on October 11, 1978. He received the Dr. Eng. degree *summa cum laude* in computer science engineering in 2005 from the University of Lecce, Italy. He is currently pursuing the Ph.D. degree in electronic and telecommunications engineering with the Department of Electronic and Telecommunications Engineering (DIET), University of Napoli Federico II, Italy. His research activities lie in the area of ad-hoc wireless networks protocol design. In particular, his current interests are focused on routing for mobile ad-hoc networks.

Giancarlo Ferraiuolo received the Laurea degree (summa cum laude) in electronic engineering from the Second University of Naples, Italy, in 2000, and the Ph.D. degree from the University of Naples Federico II, in 2004. In 2003, he spent a period as a Visiting Researcher in the Department of Electrical Engineering of Stanford University. In the period December 2004 - June 2007, he has been a researcher at the Department of Electronic and Telecommunication Engineering of the University of Naples Federico II. Now he is with AGCOM, the Italian National Regulatory Authority for Communications. His main scientific interests are in the fields of statistical image formation and of wireless network protocol and algorithm design.

Luigi Paura was born in Napoli, Italy, on February 20, 1950. He received the Dr. Eng. degree (summa cum laude) in electronic engineering in 1974 from the University of Napoli Federico II. From 1979 to 1984 he was with the Department of Electronic and Telecommunication Engineering, University of Napoli, Italy, first as an Assistant Professor and then as an Associate Professor. Since 1994, he has been a Full Professor of Telecommunications: first, with the Department of Mathematics, University of Lecce, Italy; then, with the Department of Information Engineering, Second University of Napoli; and, finally, from 1998 he has been with the Department of Electronic and Telecommunication Engineering, University of Napoli Federico II. He also held teaching positions at the University of Salerno, Italy, at the University of Sannio, Italy, and the University Parthenope, Napoli, Italy. In 1985-86 and 1991 he was a Visiting Researcher at the Signal and Image Processing Laboratory, University of California, Davis. At the present time, his research activities are mainly concerned with statistical signal processing, digital communication systems and medium access control in wireless networks.

## 1. Introduction

In the last ten years, Mobile Ad hoc NETwork (MANET) technologies have been tremendously growing. A MANET is an autonomous system of mobile nodes connected by wireless links, without any static infrastructure such as access points. Such kind of networks was introduced manly for military and emergency applications, but recently, thanks to the mesh paradigm, it can guarantee ubiquitous communication services, and it is mandatory when no cellular or other fixed infrastructures are available.

To reach a destination node located out of the coverage range of the sender node, a multi-hop communication strategy must be exploited; in such a case, each node has to cooperate with the other ones and acts as relay for packet transmission. In this scenario, the instability of the topology (link and node failures) due to node mobility and/or changes in wireless propagation conditions can frequently give rise to disconnected routes.

For such reasons, the design of an effective routing protocol for ad hoc scenarios is a challenging problem, and much research activitiy had been carried on in the last years, producing a plethora of different approaches and solutions. The proposals in [1] focus on discovering the shortest available route, according to some metrics, and all the traffic is routed over that path. This approach exhibits low tolerance against route failure events, since in such case it is necessary to stop the data transmissions until a new route will be discovered [2].

An interesting approach to gain tolerance against unreliable wireless links and node mobility is based on *multi-path routing*, in which multiple routes are proactively found. In order to effectively exploit the advantages of multi-path approaches, it is necessary to assess the performance gain reached by these strategies and, moreover, to evaluate the trade off between advantages and costs in adopting more complex multi-path solutions.

Different studies and proposals on multi-path routing have focused on heuristic methods to establish how many routes are needed and how to select them. The on-demand multi-path routing protocol in [3], which is an extension of the well-known DSR protocol [4], takes advantages of maintaining alternative disjoint routes to be utilized when the primary one fails. However, the performance benefits are evaluated only in few particular cases, regardless the tolerance against route failures. The AODV-BR [6], which is an extension of AODV [5], is analyzed by a numerical simulation analysis, which adopts the packet delivery ratio as performance metric. The same approach for performance evaluation is adopted in several works on multi-path routing, as in [7-11].

Some works have addressed the problem to analytically assess the multi-path benefits by resorting to graph theory, for both wireless sensor networks and MANETs. More specifically, in [12,13] the study is focused on a particular routing protocol, whereas in [14,15] the tolerance against route failures is evaluated with reference to the physical layer, namely in terms of network connectivity. In [16-17] the evaluation is performed for wireless sensor networks and, therefore, it assumes a hierarchical structure and the presence of a sink node. Finally, in [18-20] an analytical evaluation of multi-path routing is carried out by resorting to diversity coding.

The aim of this paper is to propose an analytical framework to evaluate the tolerance of multi-path route discovery processes against route failures, rather than to single out new multi-path routing discover processes. More specifically, with reference to MANET paradigm, we propose to resort to a theoretical approach based on graph theory. We first introduce an analytical framework based on the *terminal-pair routing reliability* (TPRR) [21] as measure of the tolerance of routing protocols against route failures. Unlike the packet delivery ratio, such a metric allows one to evaluate the robustness against the link failures, with respect to the number and the reliability of the discovered routes. In order to derive the analytical expression of the TPRR, we resort to the concept of *overlay graph*, namely the logical structure built by the *route discovery process* (RDP) of a routing protocol upon the physical network. In this way, the incomplete knowledge about the network topology that each node possesses is taken into account. Then, it is introduced an upper bound on the TPRR of any shortest-path RDP. This allows one to easily compare the performances gain of a multi-path RDP with respect to whatever shortest-path one. An algorithm for exact evaluation of routing reliability, both in numerical and symbolic form, is also provided.

The outline of the paper is the following: Section 2 introduces the network model and the assumptions, whereas Section 3 presents the analytical framework. Section 4 provides the reliability analysis and, finally, Section 5 gives the conclusions.

## 2. Network model and assumptions

In the following we introduce the network representation by resorting to the graph theory and present the main assumptions utilized in our analysis.

The nodes in the network are assumed to be reliable, while the links are failure-prone [22]. This assumption is reasonable for both static and mobile networks. In fact, in a static network, as in a sensor one, the failure of a link is due to the instability of wireless propagation conditions and to the capacity constraints, whereas in a mobile network, as in a MANET, the failure of a link is also due to the node mobility. In the following, we assume that the node mobility does not affect the reliability performance. Clearly, this assumption is realistic only when the node mobility is relatively low, since in such a case the packet delivery times are commonly smaller than those associated with topology changes [23]. The results of numerical simulations reported in Section 4.3 confirm the validity of such assumption for scenarios with moderate node mobility.

We model the network with a probabilistic direct graph:
$$G = (V, E, P) \qquad (1)$$
in which a vertex $v_i \in V$ denotes a node belonging to the network and an edge $e_{ij} \in E$ represents a communication link from node $v_i$ to node $v_j$. Each link is characterized by a failure probability $p_{ij} \in P$ (with $P$ denoting the link-failure probability matrix), which measures the probability that, at the transmission attempt time, the link is down. The edge failure events are assumed statistically independent of each other.

Given a probabilistic graph $G$, we define an overlay graph as:
$$G_O = (V, E_O, P_O) \qquad (2)$$
where $E_o \subseteq E$ and $P_o$ is the link-failure probability matrix associated with $E_o$.

Since a node *s* discovers (by means of the RDP) only a subset $E_{s,t} \subseteq E$ of the available links to reach a destination *t*, we can define the overlay graph built by the RDP upon the physical network topology as:
$$G_{s,t} = (V, E_{s,t}, P_{s,t}) \qquad (3)$$

In the following, we refer to the graph defined in (1) as the *physical graph*, which is a representation of the physical topology, while we refer to the graph defined in (3) as the overlay graph, to which we resort to measure the tolerance of a routing protocol against path failures.

As example, in Fig. 1 both the physical graph of a network and a related overlay graph for the flow (2,8) are depicted. Clearly, for each routing protocol and for each flow, the RDP defines a different overlay graph, which accounts for the features of the particular RDP as well as the network topology. Then, the overlay graph allows us to measure the effectiveness of the RDP adopted by any table based routing protocol. In fact, it allows one to assess the number of multiple paths for each flow, and moreover their *disjointness* degree (i.e. the number of disjoint links among a set of routes), enabling so to analytically evaluate the tolerance against path failures.

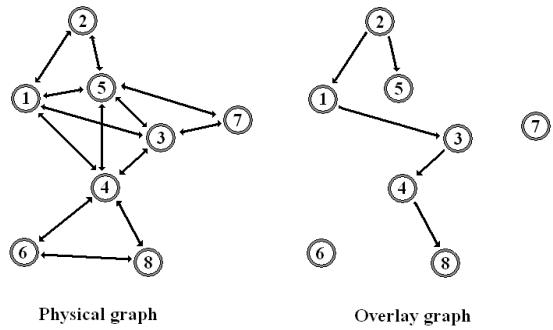

Physical graph     Overlay graph

**Figure 1 - Physical and overlay graphs**

## 3. Performance analysis framework

In this section, we present the proposed analytical framework for assessing the tolerance of RDP schemes to link failures, as well as the bound on the reliability for shortest-path RDP strategies.

*3.1 Preliminaries*

With reference to a unicast routing scenario, let us adopt as RDP performance measure the terminal-pair routing reliability (TPRR) [21], namely the probability that at least one route from the node *s* to the node *t* exists.

Considering the flow from the node s to the node t and denoting with $\Re_{s,t}$ the set of routes found by the RDP, we define the TPRR as:

$$R_{s,t}(G_{s,t}) = P(\Re_{s,t} \neq \varnothing) \qquad (4)$$

where $G_{s,t}$ is the overlay graph built by the RDP for the flow (*s,t*).

The TPRR (4) can be re-written as:

$$R_{s,t}(G_{s,t}) = 1 - \sum_{i=c}^{m} C_{s,t}(i) p^i (1-p)^{m-i} \qquad (5)$$

where $m = |E_{s,t}|$ is the cardinality of the edge set $E_{s,t}$, $p \equiv p_{i,j}$ is the link-failure probability (assumed for simplicity the same for each pair of nodes), *c* is the minimum edge cut set[1] dimension of the overlay graph between *s* and *t*, and $C_{s,t}(i)$ is the number of cut sets between *s* and *t* in the overlay graph composed exactly by *i* edges. Then, the mean TPRR is:

$$R = \frac{\sum_{s \in V} \sum_{t \in V, t \neq s} z_{s,t} R_{s,t}}{n(n-1)} \qquad (6)$$

where $n = |V|$, and $z_{s,t}$ is the probability that a data flow occurs between nodes *s* and *t*.

Accounting for the results in [24], we derive the symbolic expression of TPRR as a function of the link-failure probability *p*. More specifically, the Listing 1 shows how exactly to compute the TPRR (5) using the overlay graph. The algorithm is invoked by initializing *G* to the overlay graph $G_{s,t}$, the set *SS* to empty, and *n* to *s*. Then, the node *n* is included in the set *SS* as well as the redundant nodes, in order to ensure that the set of all emitting edges from a particular *SS* is a minimal cut set. A node is redundant if it is adjacent to *SS* and has no way to reach *t* without exploiting any node in *SS*. If the singled out set *SS* is already in the hash table *HASH*, nothing needs to be done. Differently, *SS* is a minimal cut set and it has to be added to the hash table. Then, the procedure computes the unreliability (the probability that all the links fail) for the cut set and recursively calls itself for each node adjacent to the cut set *SS*.

| Listing 1 – **Recursive(G,HASH,SS,s,t,notRel,symbNotRel)** |
|---|
| // Reliability = 1 – Recursive(…) output |
| // G is the adjacency matrix related to the overlay graph |
| // HASH is a collection of minimal cut set, initialized to empty |
| // SS is the under analysis minimal cut set, initialized to empty |
| // n is initialized to s |
| if (n == t) return; |

---

[1] An edge cut set for the flow (*s,t*) is a set of edges whose removal disconnects *s* and *t*.

```
merge(G, SS, n);                    // Merging node n in SS
absorb(G, SS, t);                   // Absorbing redundant nodes in SS
if (HASH.isPresent(SS)) return;
HASH.insert(SS);
find a cutset C of SS;
symbTempNotRel = "(1-p)^" + C.size.toString; tempNotRel = 1.0;
symbTempNotRel = symbTempNotRel + " + p * (" + symbTempNotRel;
for each edge in C
   tempNotRel = pFailed * tempNotRel;
end
for each node adjacent to SS
    Recursive(G,HASH,SS,n,t,tempNotRel,symbNotRel);
    tempNotRel = pSuccess * tempNotRel;
end
symbTempNotRel = symbTempNotRel + ")";
notRel = notRel + tempNotRel;
```

### 3.2 Polynomial bound on shortest-path reliability

In this sub-section, the performance gain achieved by a multi-path RDP with respect to any shortest-path one is estimated by resorting to an upper bound which holds for any shortest-path scheme.

The RDP of a shortest-path protocol, at best, singles out a unique route $P_{s,t}$ for the flow (*s,t*). Let us define with $h^o(s,t)$ the overlay distance between (*s,t*), i.e. the length of $P_{s,t}$ measured in number of hops on the overlay graph. Denoting with $h(s,t)$ the physical distance between (*s,t*), namely the hop distance measured on the physical graph, we have:

$$h(s,t) \leq h^o(s,t) \quad \forall s,t \in V \tag{7}$$

since the link set $E_{s,t}$ of the overlay graph is a sub-set of the link set $E$ of the physical graph and so the overlay distance $h^o(s,t)$ can not be less then $h(s,t)$.

So, the TPRR for a shortest path routing protocol can be upper bounded as:

$$R_{s,t}(G_{s,t}) = (1-p)^{h^o(s,t)} \leq (1-p)^{h(s,t)} \tag{8}$$

To estimate the distance $h(s,t)$ which clearly depends on the network topology, we make some reasonable assumptions. More specifically, we assume, according to [25], that the node density $\delta$ is uniform (according to the first interference principle) as well as the transmissions range *r*, and the physical network area *A* is a circle. Moreover, we assume the traffic pattern random as [25], namely each destination node is chosen with equal probability ($z_{s,t} \equiv z$), and the node s is at the centre of the network (to neglect the boundary effect). Under these conditions, the number of nodes in the circle of radius *x* is:

$$n(x) = \pi x^2 \delta, \quad 0 \leq x \leq \sqrt{\frac{A}{\pi}} \tag{9}$$

The probability that the node *s* communicates with a node belonging to a circular neighborhood of radius *x* can be written as:

$$P(X \leq x) = \frac{\pi x^2}{A}, \quad 0 \leq x \leq \sqrt{\frac{A}{\pi}} \tag{10}$$

where *X* is the random variable representing the path length between (*s,t*). From (10), the probability density function is:

$$f_X(x) = \frac{2\pi x}{A}, \quad 0 \leq x \leq \sqrt{\frac{A}{\pi}} \tag{11}$$

Consequentially, the average path length $\overline{L}$, measured in distance units, is:

$$\overline{L} = E[X] = \int_0^{\sqrt{A/\pi}} x f_X(x) dx = \frac{2\sqrt{A}}{3\sqrt{\pi}} \quad (12)$$

and the average physical distance, measured in number of hops, is:

$$\overline{h} = \left\lceil \frac{\overline{L}}{r} \right\rceil = \left\lceil \frac{2\sqrt{A}}{3\sqrt{\pi}r} \right\rceil = \left\lceil \frac{2\sqrt{n/\delta}}{3\sqrt{\pi}r} \right\rceil \quad (13)$$

where n is the total number of nodes in the network and $\lceil \ \rceil$ rounds to the higher integer.

Thus, the upper bound on the TPRR for any shortest path RDP is:

$$R_{s,t}(G_{s,t}) \leq (1-p)^{\left\lceil \frac{2\sqrt{n/\delta}}{3\sqrt{\pi}r} \right\rceil} \quad (14)$$

## 4 Reliability analysis

The aim of this section is twofold:
i. to show the effectiveness of the proposed analytical framework to assess the tolerance against link failures for any RDP strategy;
ii. to state performance comparisons among them.

At this end, three shortest-path routing protocols, OLSR [26], DART [27] and AODV [29], and two multi-path ones, ATR [28] and AOMDV [30], are considered. More specifically, OLSR and DART are both proactive protocols, and DART, unlike OLSR, is hierarchical, i.e. it groups the nodes belonging to the network in zones, namely siblings, and stores a unique route towards each zone for scalability purposes. AODV is a reactive routing protocol, while AOMDV generalizes AODV to exploit multiple paths with disjoint links between the source and the destination. Analogously, ATR generalizes DART, looking for multiple routes towards the same zone.

### 4.1 Overlay graph generation

The overlay graphs needed to compute the mean TPRR have been generated by simulation using Network Simulator 2 (ns-2) [31]. Fig. 2 shows the generating process of the overlay graphs.

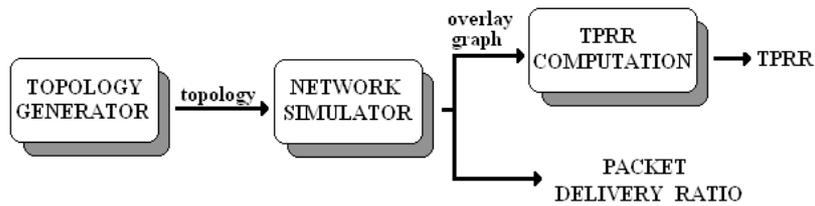

**Figure 2 - Overlay graph generating process**

For each network topology, we run a ns-2 based simulation in order to populate the routing table of each node. The paths information embedded in the routing table is then used to generate the overlay graph for each particular flow (s,t). The choice of using ns-2 to generate the routing tables has the following two advantages:

i. the overlay graphs are straight generated by the RDP utilized by the specific routing protocol;
ii. the analysis can be easily extended to different routing protocols with a light effort, simply providing to the protocol code a function which prints out the node routing table.

## *4.2 Reliability assessment*

The main characteristics of the reliability assessment setup are briefly summarized in the following. We adopt the standard ns-2 values for both the physical and the link layer to simulate an IEEE 802.11a Lucent network interface with Two-Ray Ground as propagation model. The duration of simulation is set to 500 seconds to allow the routing tables to become consistent with respect to the network topology. The sizes of the scenario areas are chosen to keep the node density equal to 64 nodes/Km2, which avoids the presence of isolated nodes [32] by assuring a mean node connectivity degree of 12. The network topologies are randomly generated by independently and uniformly distributing the nodes in the scenario area.

We have performed measures for 100 topologies for each network size. More specifically, we have reported the TPRR for the shortest-path RDPs (OLSR, DART and AODV), the shortest-path upper bound on TPRR, and the TPRR for the multi-path RDPs (AOMDV and ATR). Each figure shows the average and the variance of TPRR for each protocol as function of the link-failure probability.

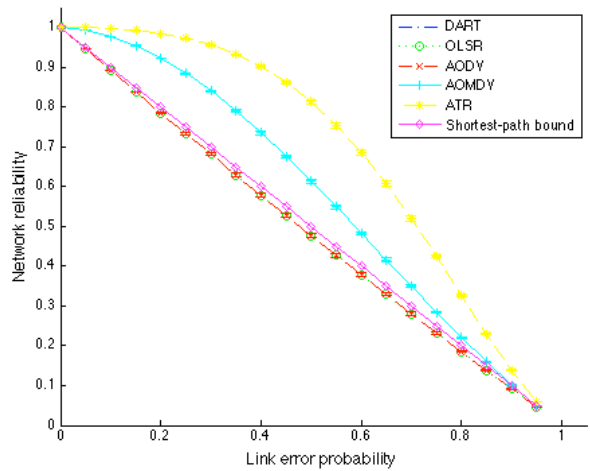

**Figure 3 – 4 nodes full mesh network**

Fig. 3 refers to a 4 nodes full-mesh network. In this case, the average TPRR reached by the shortest-path protocols agrees with the shortest-path upper bound. This means that, for very small networks, their RDP is often able to find the optimal route (one-hop route) between each pair of nodes. We note that DART RDP reaches lower values of TPRR with respect to other shortest-path RDPs, although the differences cannot be recognized in the figure. Regarding multi-path RDPs, both AOMDV and ATR outperform the shortest-path protocols also in such a small network.

*Title*

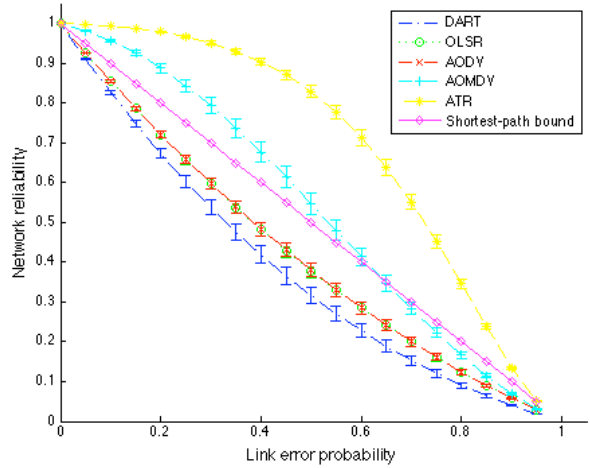

**Figure 4 – 8 nodes network**

Fig. 4 refers to a network with 8 nodes. In this case, the shortest-path protocols experience lower values of TPRR with respect to the shortest-path upper bound. Since the node connectivity degree is 12, every pair of nodes is physically linked and so the optimal route is one-hop long, in accordance with the upper bound depicted in Fig. 4. However, the shortest-path RDPs reach lower values, i.e. they discover longer routes than the optimal ones. DART RDP performs worst due to its hierarchical nature, and the largest difference is about 0.08 in correspondence of the link-failure probability *p*=0.5.

Regarding to AOMDV, for low link-failure probability, it outperforms any shortest-path protocol thanks to its multi-path characteristic, whereas, when the link-failure probability increases, such behaviour does not apply. This behaviour is reasonable, since AOMDV adopts the same route discovery of AODV, so that neither it can find the optimal routes.

Since ATR is a proactive routing protocol, it persistently broadcasts routing packets in order to discovery redundant routes. Therefore, it is able to find more paths than AOMDV. Clearly, the ATR routing overhead is higher than AOMDV one.

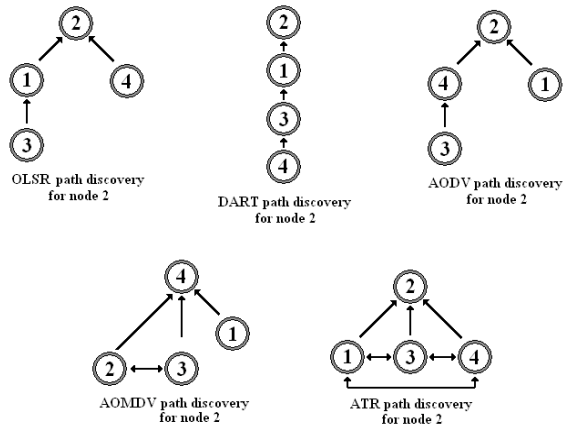

*Author*

**Figure 5 – Route discovery process**

The behaviour of shortest-path protocols depicted in Fig. 4 can be interpreted by resorting to Fig. 5, which shows an example of the routes discovered by different RDPs. The first row shows the overlay graphs built by the shortest-path RDPs towards the node '2' from three source nodes ('1', '3' and '4'). In this case, any RDP is not able to find out the optimal route for every flow and DART, due to its hierarchical nature, finds out less optimal routes than other ones. The second row presents the routes towards the node 4 singled out by the multi-path RDPs, which are able to discover redundant paths for the same flow.

In Fig. 6 we show the results for a 16 nodes network. This scenario confirms the considerations concerning Fig. 5.

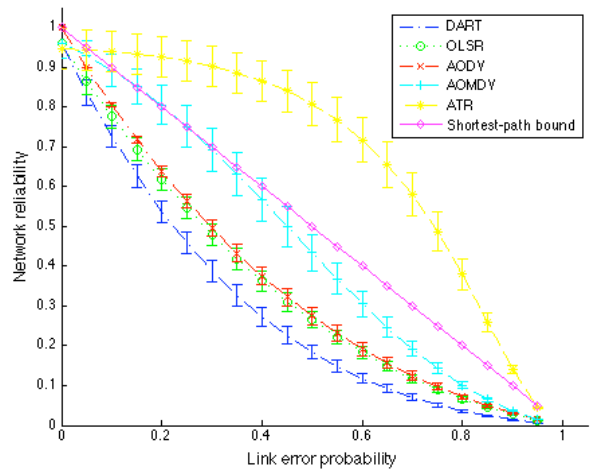

**Figure 6 – 16 nodes network**

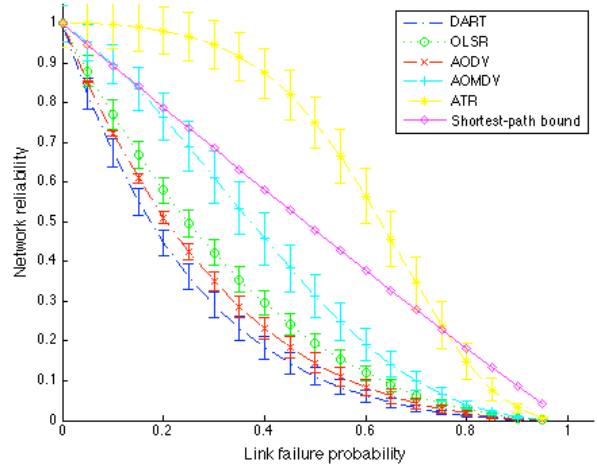

**Figure 7 – 32 nodes network**

Fig. 7 shows the results for a 32 nodes network. The AOMDV behaviour agrees with the considerations concerning the previous figures, since AOMDV RDP outperforms any shortest-path routing protocol only for low link-failure probability.

On the whole, the TPRR analysis evidences that the multi-path approach, apart from the particular RDP schema, is suitable for scenarios with *nearly reliable* links, whereas for *nearly unreliable* links the multi-path gain is negligible.

Finally, we show that the TPRR can be exploited to assess the trade-off that a routing protocol experiences between benefits due to multiple available routes and the overhead needed to discover them. In the following, we resort to TPRR to evaluate this trade-off with respect to the ATR RDP scheme.

The original ATR protocol looks for every available route towards the same zone. To analyze the mentioned trade-off, we consider two ATR RDPs which introduce a limitation in the number of discovered routes, in order to keep down the memory overhead. Specifically, in the following we analyze both the *3-limited ATR RDP* and *5-limited ATR RDP*.

Fig. 8 shows the average TPRR for a network with 16 nodes. The Fig. 8 shows that the extra overhead paid by original ATR RDP does not provide a significant performance improvement with respect to the 5-limited ATR one, which is able to exceed the upper bound on TPRR for any shortest-path RDPs for every value of *p*.

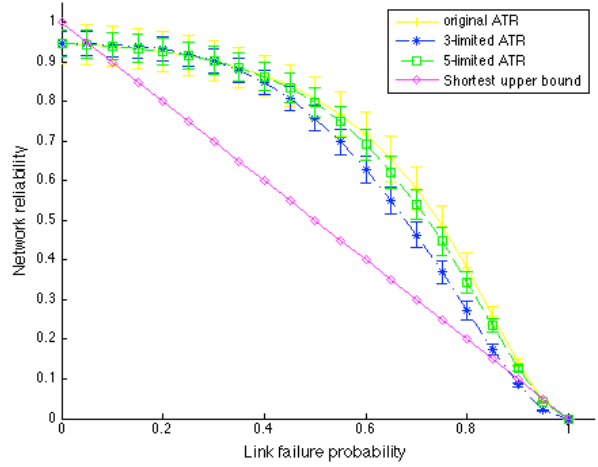

**Figure 8 – ATR RDP analysis**

## *4.3 Numerical simulations*

In this sub-section we assess the effectiveness of the proposed framework by means of a widely used routing performance metric, the *packet delivery ratio* (PDR). Clearly, the PDR is an overall metric, which measures the performance of the whole routing process, whereas the TPRR measures the only RDP performances.

The PDR measures the probability that a packet is received by the destination, while TPRR estimates the probability that at least one route exists toward the destination. It is evident that there exists dependence between the two metrics. If there is no route toward the destination the PDR has to be zero, and if all packets are correctly received than there exists at least a reliable route toward the destination. Clearly, the availability of good paths, i.e. high reliability, does not imply that the *packet forwarding algorithm* will be able to use them efficiently. Therefore, we have reported on the same figure both the TPRR and the PDR, just to verify the effectiveness of the proposed framework.

More specifically, to evaluate the PDR, we have modified both the physical and the link layer of ns-2. Regarding the former, we have introduced a uniform link-failure probability *p* for the data packets; clearly, this modification does not affect the routing and MAC packets, preserving so the RDP behaviour. Regarding the latter, we have disabled the MAC retransmission for the data packets. The duration of simulation is set to 1500 seconds. The data traffic is modelled as a CBR flow over UDP protocol with a packet rate of 1 packet/s. The start-time is at 500s end the data traffic stops at the end of the simulation. The node number is 16 and the static network topologies are the same of Section 4.2. To generate the mobile network topologies, we have adopted, as mobility model, the Random Way-Point to simulate a moderate mobility: the speed values are uniformly taken in the [0.5m/s; 5m/s] range and the pause ones in [0s, 100s].

We have performed 100 trials for each protocol and for each value of *p*. The following figures report the average TPRR, the shortest-path upper bound on TPRR and the average PDR for both static and mobile topologies, as well as the variances.

*Title*

In Fig. 9 we show the results for AODV. The PDR measured on static topologies agrees very well with the TPRR. Such a behaviour can be justified by recognizing that, in this case, the RDP is the one relevant in the overall packet delivery process. In case of mobile topologies, the behaviour is less marked. When a packet does not reach the destination, the sender starts a new route discovery. If the topology is static, the new and the previous routes will be the same, giving rise poor performances, whereas, the RDP can get the advantage by node mobility, since in such a case better routes can be discovered and used for long time intervals.

The results of Fig. 10, which refers to DART, confirm the considerations concerning Fig. 9.

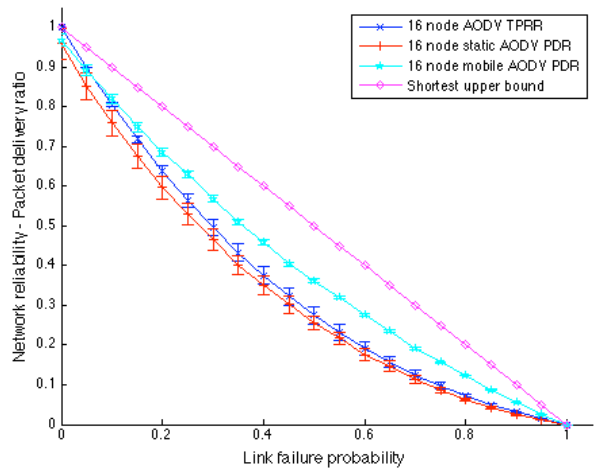

**Figure 9 – AODV PDR analysis**

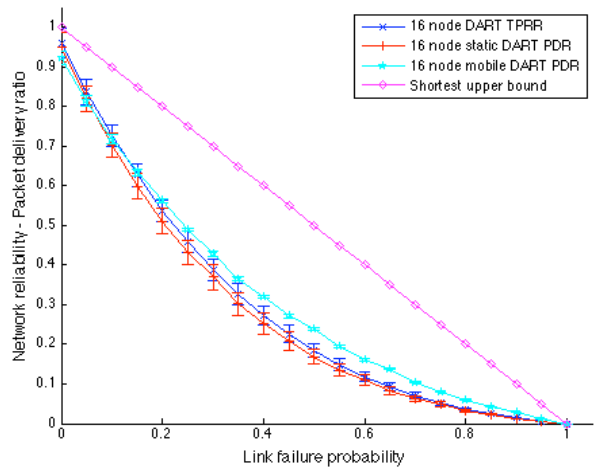

**Figure 10 – DART PDR analysis**

Fig. 11 refers to ATR; in this case, the PDR measured on static topologies does not perfectly agree with the TPRR, even if the two metrics present the same trend. We assume that the ATR *packet forwarding process*, which is liable for choosing one of the available paths, does not pick every time the best route, since it uses only local information for the selection process. The behaviour of the PDR in presence of mobility confirms the considerations concerning Fig. 9.

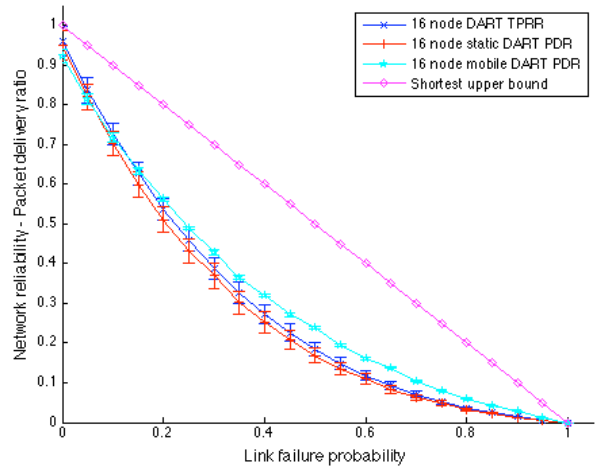

**Figure 11 – ATR PDR analysis**

## 5  Conclusion and future work

In this paper, we propose an analytical framework to evaluate the tolerance of multi-path route discovery processes against route failures, based on graph theory and on terminal pair routing reliability (TPRR) as performance measure. More specifically, it has been carried out a reliability analysis of both shortest-path and multi-path routing protocols. Resorting to numerical simulations based on a widely adopted routing performance metric, namely the packet delivery ratio, the performance results have been validated. The simulation results show the effectiveness of TPRR as performance measure.

**Acknowledgment**

This work is partially supported by Italian National project "Wireless 8O2.16 Multi-antenna mEsh Networks (WOMEN)" under grant number 2005093248 and by "Accesso Intelligente all'Informazione integrata dei BEni Culturali in ambito Regionale"(AIBER).

*Title*